\documentclass{elsart}

\usepackage{amsmath}
\usepackage{graphicx}
\usepackage{color}

\newcounter{bla}

\begin{document}
\begin{frontmatter}

\title{First-principles calculation of x-ray dichroic spectra within the full-potential linearized augmented planewave method: An implementation into the Wien2k code}

\author[a,b,d]{Lorenzo Pardini},
\author[b]{Valerio Bellini},
\author[b,c]{Franca Manghi},
\author[d]{Claudia Ambrosch-Draxl}

%\thanks[Lorenzo Pardini]{Corresponding author}
\thanks[Lorenzo Pardini] {Corresponding author.\\ \textit{E-mail address:} loren.pard@gmail.com}

\address[a]{ICCOM-CNR, Via G. Moruzzi 1, I-56124 Pisa, Italy}
\address[b]{S3 - Institute Nanoscience - CNR, Via Campi 213/A, I-41125 Modena, Italy}
\address[c]{Dipartimento di Fisica, Universit\'a di Modena e Reggio Emilia, Via Campi 213/A, I-41125 Modena, Italy}
\address[d]{Chair of Atomistic Modelling and Design of Materials, University of Leoben, Franz-Josef Stra\ss e 18, A-8700
Leoben, Austria}
%\address[b]{Second Address}

\begin{abstract}
X-ray absorption and its dependence on the polarization of light is a powerful tool to investigate the orbital and spin moments of magnetic materials and their orientation relative to crystalline axes. Here, we present a program for the calculation of dichroic spectra from first principles. We have implemented the calculation of x-ray absorption spectra for left and right circularly polarized light into the Wien2k code. In this package, spin-density functional theory is applied in an all-electron scheme that allows to describe both core and valence electrons on the same footing. The matrix elements, which define the dependence of the photo absorption cross section on the polarization of light and on the sample magnetization, are computed within the dipole approximation. Results are presented for the $L_{2,3}$ and $M_{4,5}$ egdes of CeFe$_{2}$ and compared to experiments.

\begin{keyword}
first-principles; DFT; XMCD; full-potential LAPW; Wien2k.
  % Please give some freely chosen keywords that we can use in a
  % cumulative keyword index.
\end{keyword}

\end{abstract}

\end{frontmatter}

\newpage

% In program descriptions the main text of the paper is listed under
% the heading LONG WRITE-UP.

\hspace{1pc}
%{\bf LONG WRITE-UP}

\section{Introduction}

Dichroism is the property of a material to absorb photon beams of different polarization with different cross-sections. In particular, x-ray magnetic circular dichroism (XMCD) originates from excitations of core electrons to unoccupied spin-split conduction states, giving rise to different absorption behavior of left and right polarized light. XMCD, together with x-ray absorption spectroscopy (XAS), is one of the most effective tools for obtaining information about magnetic systems. It exhibts different capabilities of characterizing these systems which cannot be afforded by traditional magnetic techniques. Among the advantages of XMCD, it is worth mentioning its high sensitivity and chemical selectivity, the latter being an essential property to study magnetism in alloys, oxides, impurities, surfaces, and interfaces. Moreover, it represents one of the few methods which can discriminate between orbital and spin contribution to the total magnetic moment. Information about spin and orbital magnetic moments, associated to the photo-absorbing atom are extracted from XMCD spectra, exploiting two specific sum rules \cite{thole1992, carra1993} which allow to extract the expectation value of the orbital and spin angular momentum operators, $L_{z}$ and $S_{z}$.

We have developed a package for the calculation of x-ray absorption spectra at the K-, L-, and M-edges for left and right circularly polarized light in the Wien2k code \cite{wien2k}.\footnote{XAS calculations can already be performed in the Wien2k distribution via the \emph{XSPEC} package, but in that implementation the polarization of light is not considered, thus dichroic spectra can not be trivially derived.} More specific, we have extended the Wien2k {\tt OPTIC} package \cite{claudia2006}, which allows to calculate optical properties. The matrix elements, which define the dependence of the photo-absorption cross-section on the polarization of light and on the sample magnetization, are calculated within the dipole approximation.

In the following, we will give an extensive description of the formulae behind the method. As prototypical examples, we have investigated dichroic and absorption spectra at the $M_{4,5}$ and $L_{2,3}$ edges of cerium in $CeFe_2$, which demonstrate the applicability of our code to such core excitations.

\section{Theoretical background}
\label{xmcdBasic}

XMCD is determined by the difference in absorption between left and right circularly polarized light, where left and right are referred to the propagation direction of incoming radiation with respect to the magnetization axis of the system. In particular, photons are left (right) polarized or, equivalently, their helicity is $-\hbar$ ($+\hbar$), when the direction of propagation is anti-parallel (parallel) with respect to the magnetization.

The absorption of a photon with polarization vector $\hat{\epsilon}$ results in the excitation of a selected atom from an initial state $\vert \Psi_{i} \rangle$ into a final state $\vert \Psi_{f} \rangle$. In the framework of the electric-dipole approximation, the cross section $\mu(\omega)$ for x-ray absorption is given by Fermi's golden
rule:
\begin{equation}
\label{eq3} \mu(\omega) \propto \sum_{f} | \hat{\epsilon} \cdot
\textbf{D}_{f i}|^{2} \delta (E_{f}-E_{i}- \hbar \omega),
\end{equation}
where
\begin{equation}
 \label{eqc1_1}
\textbf{D}_{f i}=<\Psi_{f}\vert  \textbf{p}\vert\Psi_{i}>
\end{equation}
is the matrix element of the
%moment $\textbf{p} = (\imath \hbar )^{-1} \nabla_{\textbf{r}} $
momentum operator $\textbf{p} = -\imath \hbar \nabla_{\textbf{r}}$
between the initial and final states of energy $E_i$ and $E_f$. It is convenient to express the operator $ \nabla$  in terms of \textit{spherical components}
\begin{equation}
 \label{eqc1_2} \nabla_{\pm 1} = \mp\frac{1}{\sqrt{2}}\left( \frac{\partial}{\partial_{x}} \pm \imath\frac{\partial}{\partial_{y}}\right) ; \ \ \ \  \nabla_{0} = \frac{\partial}{\partial_{z}}.
\end{equation}
Similarly, the polarization vector for right ($+1$) and left ($-1$) polarized light is
\begin{equation}
 \label{eqc1_4} \epsilon_{\pm 1}=   \epsilon_x \pm  \imath  \epsilon_y
\end{equation}
and the scalar product appearing in Eq. \ref{eq3} will select the $\pm 1$ components of the vector $\textbf{D}_{f i}$. The absorption cross section for the two polarizations then becomes
\begin{equation}
\label{eqpm} \mu^{\pm}(\omega) \propto \sum_{f} | D^{\pm 1}_{fi}|^{2} \delta (E_{f}-E_{i}- \hbar \omega),
\end{equation}
where
\begin{equation}
\label{momentummcore}
D^{\pm 1}_{fi}=\epsilon^{\pm}\cdot
\langle \Psi_{i}\vert \textbf{p} \vert
\Psi_{f} \rangle= \langle
\Psi_{i}\vert \nabla_{\pm}
\vert \Psi_{f} \rangle.
\end{equation}

\subsection{Sum rules}
\label{sumrules}

A connection between the integrated XMCD signal and the ground state expectation value of the projection of orbital angular momentum on the magnetization axis was suggested by Thole, Carra, and van der Laan \cite{thole1992}. Later, the same authors derived a second sum rule for the spin moment using graphical angular momentum techniques \cite{carra1993}. They developed these sum rules by analyzing near-edge x-ray circular dichroism and using a single-ion model with a partially filled valence shell. Starting from the absorption cross section for pure dipole transitions, Eq. \ref{eq3}, integrating over the photon energy and applying the Wigner-Racah coupling techniques \cite{brouder1990, lindgren, brink, cowan}, they obtained the following relations:
\begin{equation}
\label{eq_orbgen}
\frac{\int_{j_{\pm}}d\omega(\mu^{+}-\mu^{-})}{\int_{j_{\pm}}d\omega(\mu^{+}+\mu^{-}+\mu^{0})}=\frac{1}{2}\frac{c(c+1)-\ell(\ell+1)-2}
{\ell(\ell+1)(4\ell+2-n)}\langle L_{z} \rangle
\end{equation}
\begin{eqnarray}
\label{eq_spingen}
\begin{split}
 & \frac{\int_{j_{+}}d\omega(\mu^{+}-\mu^{-})-[(c+1)c]\int_{j_{-}}d\omega(\mu^{+}-\mu^{-})}
{\int_{j_{\pm}}d\omega(\mu^{+}+\mu^{-}+\mu^{0})} = \\
 & =\frac{\ell(\ell+1)-2-c(c+1)}{3c(4\ell+2-n)}\langle S_{z} \rangle + \\
 & +\frac{\ell(\ell+1)[\ell(\ell+1)+2c(c+1)+4]-3(c-1)^{2}(c+2)^{2}}{6\ell c(\ell+1)(4\ell+2-n)}\langle T_{z} \rangle ,
\end{split}
\end{eqnarray}
where $c$ represents the core-hole orbital quantum number, $n$ the number of electrons in the valence shell, $\ell$ the valence orbital quantum number, $j_{\pm}=c \pm 1/2 $ the quantum number of the two partners of the spin-orbit split inner shell, $\mu_{0}=(\mu_{+}+\mu_{-})/2$ the absorption cross section for incident light polarized along the direction of magnetization, and $T_{z}$ the magnetic dipole operator defined
as:
\begin{equation}
\label{eq_tz}
T_{z}=\left[ \sum_{i}\textbf{s}_{i}-3\textbf{r}_{i}(\textbf{r}_{i}\cdot \textbf{s}_{i})  \right]_{z}.
\end{equation}

\section{X-ray circular dichroism within the LAPW basis set}
\label{theory}

\subsection{Density functional theory and the LAPW basis set}
\label{DFT}

In DFT-based band structure calculations \cite{HK}, one-electron wavefunctions, $\Psi_{n}({\bf r})$, and eigenvalues, $E_{n}$, are derived solving the Kohn-Sham(KS) equations \cite{KS}:
\begin{equation}
\left[ -\nabla^{2} +V_{H}(\textbf{r})+V_{nucl}(\textbf{r})
+V_{xc}(\textbf{r}) \right]
\Psi_{n}({\bf r}) = E_n \Psi_{n}({\bf r})
\label{KS-equation}
\end{equation}
where $V_{H}(\textbf{r})$ denotes the Hartree potential, $V_{nucl}(\textbf{r})$ is the bare Coulomb potential of the atomic nuclei, and $V_{xc}(\textbf{r})$ is the exchange-correlation potential, which is defined as the functional derivative of the energy $E_{xc}[n]$, n(r) being the ground-state density of the system. In order to solve these equations numerically, the KS orbitals are expanded in terms of an appropriate finite set of basis functions $\left\{ \phi_{\nu}\right\}$,
\begin{equation}
\Psi_n({\bf r})=\sum_{\nu}C_{\nu}^n \phi_{\nu}({\bf r}),
\label{ansatz}
\end{equation}
and the coefficients $C_{\nu}^n$ are then obtained by diagonalizing the Hamiltonian matrix.

If one is interested in exploring the physics of the core region, as in the case of the XMCD, all-electron schemes are needed. Among them, the ones descending from the augmented planewave (APW) method are the most precise ones for a proper description of magnetic properties. These methods employ a hybrid set of basis functions, \emph{i.e.}, atomic-like basis functions in the unit-cell region close to nuclei and planewaves elsewhere. Our method of choice is the linearized augmented planewave (LAPW) method \cite{andersen1975,koelling1975,singh1994,cad2004}, where the basis functions inside the {\it muffin-tin spheres} are linear combinations of the radial functions $u_{\ell}^{\alpha }(r,E_\ell)$ and their energy derivatives ${\dot u}_{\ell}^{\alpha }(r,E_\ell)$ at the trial energy $E_\ell$ times spherical harmonics $Y_{\ell m}({\bf{{\hat r}}})$.
%
%\begin{widetext}
\begin{equation}
\phi_{{\bf k}+{\bf G}}  ({\bf S}_{\alpha} + {\bf r}) =
\sum_{\ell m}[A_{\ell m}^{\alpha}({\bf k} + {\bf G}) u_{\ell}^{\alpha}(r,E_\ell) +
B_{\ell m}^{\alpha}({\bf k} + {\bf G}) \dot{u}_{\ell}^{\alpha}(r,E_\ell)]
Y_{\ell m}({\bf{{\hat r}}})
\label{basis_LAPW}
\end{equation}
%\end{widetext}
%
with \textbf{G} denoting a reciprocal space vector. The radial functions are obtained by solving the radial Schr\"odinger equation in the spherical potential of the respective atomic sphere. The coefficients $A_{\ell m}^{\alpha }({\bf k}+{\bf G})$ and $B_{\ell m}^{\alpha }({\bf k}+{\bf G})$ are determined for each atom by matching the two types of basis functions at the atomic sphere boundary, both in value and slope. Fixing the energies $E_\ell$ in Eq. \ref{basis_LAPW} makes the basis set energy independent. As a consequence, the secular equation becomes linear in energy, leading to a generalized eigenvalue problem. But, at the same time, one faces the restriction that for a given $\ell$ value only the sates of one principal quantum number can be described. A solution of this issue is the introduction of additional basis functions, called {\it Local Orbitals} \cite{singh1991b} (LO) of the form
%
%\begin{widetext}
\begin{equation}
\phi_{LO}  ({\bf S}_{\alpha} + {\bf r}) =
\left[
{\tilde A}_{\ell m}^{\alpha} u_{\ell}^{\alpha}(r,E_\ell) +
{\tilde B}_{\ell m}^{\alpha} \dot{u}_{\ell}^{\alpha}(r,E_\ell) +
{\tilde C}_{\ell m}^{\alpha} u_{\ell}^{\alpha}(r,E_{lo})
\right]
Y_{\ell m}({\bf{{\hat r}}})
\label{basis_LO}
\end{equation}
%\end{widetext}
%
where $E_{\ell}$ is the same as in the LAPW basis (Eq. \ref{basis_LAPW}), and $E_{lo}$ represents the trial energy of the semicore state. Here ${\tilde A}_{\ell m}^{\alpha}$ and ${\tilde B}_{\ell m}^{\alpha}$ are determined such that the LO and its slope go to zero at the sphere boundary, {\it i.e.}, these basis functions are completely confined within the atomic spheres, whereas ${\tilde C}_{\ell m}^{\alpha}$ are chosen in order to normalize the basis function.

An alternative to the linearization described above is provided by the APW+\lowercase{lo} method \cite{sjostedt2000}. Here, the corresponding basis consists of APW functions taken at a fixed energy $E_\ell$
\begin{equation}
\phi_{{\bf k}+{\bf G}}  ({\bf S}_{\alpha} + {\bf r}) =
\sum_{\ell m}A_{\ell m}^{\alpha}({\bf k} + {\bf G}) u_{\ell}^{\alpha}(r,E_\ell)
Y_{\ell m}({\bf{{\hat r}}})
\label{basis_APW+lo}
\end{equation}
supplemented by local orbitals of the form
\begin{equation}
\phi_{lo}  ({\bf S}_{\alpha} + {\bf r}) =
\left[
{\tilde A}_{\ell m}^{\alpha,lo}           u_{\ell}^{\alpha}(r,E_\ell) +
{\tilde C}_{\ell m}^{\alpha,lo}  {\dot u}_{\ell}^{\alpha}(r,E_\ell)
\right]
Y_{\ell m}({\bf{{\hat r}}}).
\label{basis_lo}
\end{equation}
The two coefficients $A_{\ell m}^{\alpha,lo}$ and $C_{\ell m}^{\alpha,lo}$ are determined by normalization and by requiring local orbitals to have zero value at the muffin-tin boundary, but not zero slope.

The Wien2k code can supply both types of basis functions, \emph{i.e.}, APW+lo and LAPW+LO, and it is possible to choose one or the other independently for different atoms and angular momentum numbers.

\subsection{The XMCD formalism in the LAPW basis}
\label{implementation}

As we have seen in Sec. \ref{xmcdBasic}, the main ingredient for the calculation of a dichroic signal is the expression of the x-ray absorption yield. Within the one-particle framework and in the dipole approximation, the absorption cross-section $\mu$ for incident x-rays is determined by the probability of an electron to be excited from a core state $\Psi_{j,m_{j}}(\textbf{r})$ with energy $E_{j}$, to a final valence state $\Psi_{n}^{\textbf{k}}(\textbf{r})$ with energy $E_{f}^{\textbf{k}}$, according to Fermi's golden rule (Eq. \ref{eq3}). Note that core states are characterized by atomic quantum numbers $j,m_{j}$ and valence states by band index $f$ and momentum  $\textbf{k}$. From this formula, dichroic and total absorption signals can be calculated through the linear combinations $\mu^{+}-\mu^{-}$ and $\mu^{+}+\mu^{-}$, respectively. Therefore, the main task is to compute the momentum matrix elements.

In Wien2k, core states are calculated by the routine \emph{LCORE}, which represents a modified version of the relativistic atomic LSDA code by Desclaux \cite{desclaux1969, desclaux1975} that solves the fully relativistic Dirac equation. Core states wave function can thus be written as:
\begin{equation}
\label{eq_core}
\Psi_{j,m_{j}}(\textbf{r}) = \sum_{M=-L}^{L} \sum_{m_{s}=-1/2}^{1/2} C^{jm_{j}}_{ML,s m_{s}}\varphi^{c}_{ML}(\textbf{r}),
\end{equation}
where
\begin{equation}
\label{eq_corebasis}
\varphi^{c}_{LM}=u_{L}^{c}(r)Y_{LM}(\hat{r})
\end{equation}
and $u_{L}^{c}(r)$ are the solutions of the radial part of the Dirac equation.
Valence states are expanded in terms of the LAPW basis set according to Eq. \ref{ansatz}, here rewritten as:
\begin{equation}
\Psi^{\bf k}_{f}({\bf r})=\sum_{\textbf{G}}C_{f}^{\bf k}(\textbf{G})\phi_{\textbf{k}+\textbf{G}}({\bf r})
\label{ansatz2}
\end{equation}
Inserting Eqs. \ref{eq_core} and \ref{ansatz2} in Eq. \ref{momentummcore}, we obtain
\begin{eqnarray}
\label{eq_mme3}
\textbf{D}^{\pm}_{f,j,m_{j}}(\textbf{k})&=&\langle \Psi_{j,m_{j}}\vert  \nabla_{\pm} \vert
\Psi_{f}^{\textbf{k}}\rangle= \nonumber \\
 &=& \sum_{\textbf{G}} \sum_{M} C_{f}^{\textbf{k}}(\textbf{G}) \langle \sum_{m_{s}} C^{jm_{j}}_{LM, s m_{s}}
\varphi^{c}_{L M}(\textbf{r})\vert
 \nabla_{\pm} \vert \phi_{\textbf{k}+\textbf{G}}(\textbf{r}) \rangle.
\end{eqnarray}
These matrix elements must be inserted into Eq. \ref{eqpm} which, after summing over $m_j$, becomes:
\begin{equation}
\label{eqpm2} \mu^{\pm}(\omega) \propto \sum_{f} \sum_{m_j=-j}^{j} | \textbf{D}^{\pm}_{f,j,m_j}(\textbf{k})|^{2}
\delta (E_{f}^{k}-E_{j}- \hbar \omega)
\end{equation}
The $\nabla_{\pm} $ operators can be expressed in spherical coordinates as
\begin{equation}
\label{eq_sfer}
 \nabla_{\pm}=sin \theta e^{\pm i\phi}\frac{\partial}{\partial r}+\frac{1}{r}
e^{\pm i\phi} \left( cos\theta \frac{\partial}{\partial \theta} \pm \frac{i}{sin \theta} \frac{\partial}{\partial \phi} \right).
\end{equation}
In the case of positive helicity of the photon, the $ \nabla_{+}$ operator applied
to the valence state gives:
\begin{equation}
\label{eq_popui}
    \begin{split}
&\left(  \nabla_{+} \right) \phi_{\textbf{k}+\textbf{G}}(\textbf{r}) =  \\
&=\sum_{\ell m} \frac{\partial}{\partial r} \left[ A^{\alpha}_{\ell m}(\textbf{k}+\textbf{G})u^{\alpha}_{\ell}(r,E_{\ell})
+B^{\alpha}_{\ell m}(\textbf{k}+\textbf{G})\dot{u}^{\alpha}_{\ell}(r,E_{\ell}) \right] sin\theta e^{i\phi}Y_{\ell m}(\hat{\textbf{r}})+  \\
&+\frac{1}{r}\sum_{\ell m} \left[ A^{\alpha}_{\ell m}(\textbf{k}+\textbf{G})u^{\alpha}_{\ell}(r,E_{\ell})
+B^{\alpha}_{\ell m}(\textbf{k}+\textbf{G})\dot{u}^{\alpha}_{\ell}(r,E_{\ell}) \right]  \cdot   \\
&\cdot e^{i\phi}
\left( cos \theta \frac{\partial}{\partial \theta} +i \frac{i}{sin\theta}\frac{\partial}{\partial\phi} \right) Y_{\ell m}(\hat{\textbf{r}})
    \end{split}
\end{equation}
Exploiting the following relations between spherical harmonics:
\begin{equation}
\label{eq_shperpropp}
e^{+i\phi}sin\theta Y_{\ell m}=F^{(1)}_{\ell m} Y_{\ell +1,m+1}+F^{(2)}_{\ell m} Y_{\ell -1,m+1}\\
\end{equation}
\begin{equation}
\label{eq_shperpropm}
e^{-i\phi}sin\theta Y_{\ell m}=F^{(3)}_{\ell m} Y_{\ell +1,m-1}+F^{(4)}_{\ell m} Y_{\ell -1,m-1}\\
\end{equation}
\begin{equation}
\label{eq_shperproppp}
e^{+i\phi} \left( cos\theta \frac{\partial}{\partial \theta}+\frac{i}{\sin \theta} \frac{\partial}{\partial \phi} \right)
Y_{\ell m}=-\ell F^{(1)}_{\ell m}Y_{\ell +1,m+1}+(\ell +1)F^{(2)}_{\ell m} Y_{\ell -1,m+1}
\end{equation}
with
\begin{equation}
\label{eq_effe1}
F^{(1)}_{\ell m}=-\sqrt{\frac{(\ell +m+1)(\ell +m +2)}{(2\ell +1)(2\ell +3)}}\\
\end{equation}
\begin{equation}
\label{eq_effe2}
F^{(2)}_{\ell m}=\sqrt{\frac{(\ell -m)(\ell -m -1)}{(2\ell -1)(2\ell +1)}}\\
\end{equation}
\begin{equation}
\label{eq_effe11}
F^{(3)}_{\ell m}=\sqrt{\frac{(\ell -m+1)(\ell -m +2)}{(2\ell +1)(2\ell +3)}}\\
\end{equation}
\begin{equation}
\label{eq_effe22}
F^{(4)}_{\ell m}=-\sqrt{\frac{(\ell +m)(\ell +m -1)}{(2\ell -1)(2\ell +1)}}\\
\end{equation}
the momentum matrix element relative to right polarized radiation becomes:
\begin{equation}
\label{eq_mme4}
\begin{split}
&\textbf{D}^{+}_{f,j,m_{j}}(\textbf{k}) = \langle \Psi_{j,m_{j}} \vert  \nabla_{+} \vert \Psi_{f}^{\textbf{k}} \rangle = \\
 &=\sum_{\textbf{G}} \sum_{M} \sum_{\ell m} C_{f}^{\textbf{k}} (\textbf{G}) \int r^{2} dr \int d\Omega
\sum_{m_{s}} C^{jm_{j}}_{L M, s m_{s}} u^{c}_{L}(r)Y^{*}_{LM}(\hat{\textbf{r}}) \cdot \\
 &\cdot \sum_{m \ell} \{ [ \underbrace{A_{\ell m} u'_{\ell} +B_{\ell m} \dot{u}'_{\ell}}_{W'_{\ell m}}-
\frac{\ell}{r}\underbrace{(A_{\ell m} u_{\ell} +B_{\ell m} \dot{u}_{\ell})}_{W_{\ell m}}]F^{(1)}_{\ell m}Y_{\ell +1,m+1}+ \\
 &+[\underbrace{A_{\ell m} u'_{\ell} +B_{\ell m} \dot{u}'_{\ell}}_{W'_{\ell m}}+
\frac{\ell +1}{r}\underbrace{(A_{\ell m} u_{\ell} +B_{\ell m} \dot{u}_{\ell})}_{W_{\ell m}}]F^{(2)}_{\ell m}Y_{\ell -1,m+1} \}=\\
 &=\sum_{\textbf{G}} \sum_{M} \sum_{\ell m} \sum_{m_{s}} C^{jm_{j}}_{L M,s m_{s}} C_{f}^{\textbf{k}}(\textbf{G}) \{ \int r^{2}
dr u^{c}_{L} [W_{\ell m}' -\frac{\ell}{r} W_{\ell m}]F^{(1)}_{\ell m}
\underbrace{\int d\Omega Y^{*}_{LM} Y_{\ell +1,m+1}}_{\delta_{L,\ell +1}\delta{M,m+1}}+ \\
 &+\int r^{2} dr u^{c}_{L} [W_{\ell m}' +\frac{\ell +1}{r} W_{\ell m}]F^{(2)}_{\ell m}
\underbrace{\int d\Omega Y^{*}_{LM} Y_{\ell -1,m+1}}_{\delta_{L,\ell -1}\delta_{M,m+1}} \}= \\
 &=\sum_{\textbf{G}} C_{f}^{\textbf{k}}(\textbf{G}) \sum_{M=-L}^{L} \sum_{m_{s}=-1/2}^{1/2} C^{jm_{j}}_{LM, s m_{s}} \\
 &\{A_{L-1,M-1} [\int u^{c}_{L}u'_{L-1}r^{2}dr-(L-1)\int u^{c}_{L}u_{L-1}rdr ]F^{(1)}_{L-1,M-1}+ \\
 &+B_{L-1,M-1} [\int u^{c}_{L} \dot{u}'_{L-1}r^{2}dr-(L-1)\int u^{c}_{L}\dot{u}_{L-1}rdr] F^{(1)}_{L-1,M-1}+ \\
 &+A_{L+1,M-1} [\int u^{c}_{L}u'_{L+1}r^{2}dr+(L+2)\int u^{c}_{L}u_{L+1}rdr ]F^{(2)}_{L+1,M-1}+ \\
 &+B_{L+1,M-1} [\int u^{c}_{L} \dot{u}'_{L+1}r^{2}dr+(L+2)\int u^{c}_{L}\dot{u}_{L+1}rdr] F^{(2)}_{L+1,M-1}
\}
\end{split}
\end{equation}
The corresponding momentum matrix element related to the $ \nabla_{-}$ operator is derived in the same way, and the following formula is obtained:
\begin{equation}
\label{eq_mme7}
\begin{split}
 &\textbf{D}^{-}_{f,j,m_{j}}(\textbf{k}) =
\sum_{\textbf{G}} C_{f}^{\textbf{k}}(\textbf{G}) \sum_{M=-L}^{L} \sum_{m_{s}=-1/2}^{1/2} C^{jm_{j}}_{LM, s m_{s}} \\
 &\{A_{L-1,M+1} [\int u^{c}_{L}u'_{L-1}r^{2}dr-(L-1)\int u^{c}_{L}u_{L-1}rdr ]F^{(3)}_{L-1,M+1}+\\
 &+B_{L-1,M+1} [\int u^{c}_{L} \dot{u}'_{L-1}r^{2}dr-(L-1)\int u^{c}_{L}\dot{u}_{L-1}rdr] F^{(3)}_{L-1,M+1}+\\
 &+A_{L+1,M+1} [\int u^{c}_{L}u'_{L+1}r^{2}dr+(L+2)\int u^{c}_{L}u_{L+1}rdr ]F^{(4)}_{L+1,M+1}+ \\
 &+B_{L+1,M+1} [\int u^{c}_{L} \dot{u}'_{L+1}r^{2}dr+(L+2)\int u^{c}_{L}\dot{u}_{L+1}rdr] F^{(4)}_{L+1,M-1}
\}
\end{split}
\end{equation}
Finally, the momentum matrix elements, Eqs. \ref{eq_mme4} and \ref{eq_mme7}, have to be inserted in Eq. \ref{eqpm2}, and the absorption spectra are then calculated for the two polarizations by summing over band indices and momenta. The total absorption and dichroic signals are derived as sum and difference of $\mu_+$ and $\mu_-$, respectively.

A few more words should be spent to see what happens when Local Orbitals are taken into account. It is worth noticing that, in this case, the basis set is augmented by a term given by Eq.\ref{basis_LO}. The Kohn-Sham wavefunction (Eq. \ref{ansatz2}) thus becomes
\begin{equation}
\label{eq_implLOLO}
\Psi_{f}^{\textbf{k}}(\textbf{r})=\sum_{\textbf{G}}C_{f}^{\textbf{k}}(\textbf{G})\phi_{\textbf{k}+\textbf{G}}(\textbf{r})
+\sum_{\ell m}C^{\textbf{k},LO}_{f,\ell m} (\textbf{G})\phi_{LO}(\textbf{r}).
\end{equation}
This means that a further term must be added to the expression for the matrix elements (Eqs. \ref{eq_mme4} and \ref{eq_mme7}). As, however, the derivation is along the same lines as above, it will not be explicitely described. The same reasoning can be applied, \emph{mutatis mutandis}, if the APW+lo basis set (Eq. \ref{basis_lo}) is used.

If we assume  the core-hole and the excited photoelectron to have both an infinite lifetime, the initial and final energies of the electronic transition are sharp, resulting in rapidly varying spectra. In reality, however,  initial and final states have finite lifetimes: The core-hole decays by radiative or Auger electronic transitions, whereas the excited electron can lose energy by emitting plasmons or creating electron-hole pairs. These finite lifetimes can be accounted for by a smearing of the spectra, \emph{i.e.}, adopting a Lorentzian broadening for the initial and final states. Usually, the (smaller) energy-dependent photoexcited electron broadening can be safely neglected, and only the more effective core-hole broadening ($\Gamma_{c}$) is considered. The broadened spectra, $\overline{F}(E)$, is obtained by $F(E)$, the unbroadened one, by:
\begin{equation}
\label{eq_corebroad}
\overline{F}(E)=\frac{\Gamma_{c}}{2\pi}\int^{+\infty}_{-\infty}
\frac{F(E')dE'}{(E-E')^{2}+\frac{1}{4}\Gamma_{c}^{2}}
\end{equation}
Particular care must be taken in choosing the broadening parameters. In the case of the $L_{2,3}$ edge, for instance, different values must be used for the two edges because, in the experiments, the $L_{2}$ line is wider than $L_{3}$. This different broadening is determined by the fact that the decay of a $p_{1/2}$ core-hole has a channel (the \emph{super-Coster-Kronig} process $p_{1/2} \rightarrow p_{3/2}$) which is not available to a $p_{3/2}$ core-hole. Finally, the spectra can be convoluted with a Gaussian function in order to take into account the finite spectrometer resolution.

\section{Implementation into the Wien2k code}
\label{WienXMCD}

The application of our package is based on a self-consistent calculation for a spin-polarized system, including spin-orbit coupling. It concerns two programs, which are {\tt OPTIC} and {\tt JOINT}, that must be run consequently. The former performs the calculation of the momentum matrix elements $D^{\pm}_{f,j,m_{j}}(\textbf{k})$ (Eq. \ref{eq_mme3}), whereas JOINT carries out the integration over the Brillouin zone (BZ) by means of Bloechl's tetrahedron method \cite{blochl1994}.

{\tt OPTIC} reads the core and valence wavefunction, which are calculated by the routines {\tt LCORE} and {\tt LAPW1}, from the files {\tt case.vectorsoup/dn} and {\tt case.corewfup/dn}, respectively. In order to perform the XMCD calculation, the user has to define the atom for which the spectra should be calculated as well as the kind of transition. These parameters must be provided in the input file {\tt case.inop}, which is described in Table \ref{tab:WienWrapOp}.

\begin{table}[h]
\caption { A typical input file {\tt case.inop} for computing the matrix elements which determine the dichroic spectra. The values correspond to the calculation of the Ce L$_{2,3}$ edge, shown in Sec. \ref{cefe2}.}
\vskip 0.2cm
\begin{tabular}{l l l } \hline
\hline
99999  \quad 1 & {\tt KUPLIMIT}, {\tt KFIRST}\\
 0.0 \quad3.2 &  {\tt EMIN}, {\tt EMAX}\\
 1 \quad 1 \quad L23  & {\tt XMCD}, {\tt ATOM\_NUM}, {\tt EDGE}\\
\hline
\end{tabular}
\label{tab:WienWrapOp}
\end{table}

The meaning of the input parameters is the following:
\begin{itemize}
  \item {\tt KUPLIMIT} is the maximum number of k-points to be taken into account; {\tt KFIRST} is the first k-point. (This value should be only different from 1 for special analysis purposes.)
  \item {\tt EMIN} and {\tt EMAX} define the absolute energy range (in Ry) in which the matrix elements are calculated.
  \item {\tt XMCD} is a
%logical
parameter to decide whether to perform an XMCD calculation ({\tt XMCD = 1}) or optical properties with the original version of the OPTIC program ({\tt XMCD = 0}).
  \item {\tt ATOM\_NUM} specifies the atom (according to {\tt case.struct} file) for which an XMCD calculation is performed.
  \item {\tt EDGE} indicates the particular edge; possible values are (case sensitive): K1, L1, L23, M1, M23, and M45.
\end{itemize}

The most important routines of the OPTIC program, which are relavant for the XMCD calculation are listed below together with a short description:

\begin{itemize}
\item {\tt opmain.f} is the main routine, reads the input parameters from the file {\tt case.inop} and the structural data from {\tt case.struct} and calls {\tt sph-UPcor.frc}.
\item {\tt sph-UPcor.frc} is the XMCD master routine. It is responsible for the computation of the momentum matrix elements according to Eq. \ref{eq_mme3}.
\item {\tt momradintc.f} calculates the radial integrals that appear in Eqs. \ref{eq_mme4} and \ref{eq_mme7}.
\item {\tt mmatcv.f} performs the summation of the radial integrals over $M$ (Eqs. \ref{eq_mme4} and \ref{eq_mme7}).
\item {\tt outmatABZ.f} calculates the squared matrix elements and sums over the the quantum number m$_j$ (Eq. \ref{eqpm2}).
\end{itemize}

The momentum matrix elements between the selected core state and the conduction states are stored in {\tt case.symmat1up} for the upper spin-orbit split core state (\emph{e.g.} L$_3$) and in {\tt case.symmat2up} for the lower spinorbit-split core state (\emph{e.g.} L$_2$) for each k-point (\textbf{k}) and every band (f). In the case of K, L$_1$, and M$_1$ edges, only {\tt case.symmat1up} is written.\footnote{Core states are treated fully relativistically. As a consequence, those states with principal quantum number $\ell$ different from zero are spin-orbit split, resulting in a state whose total angular momentum $J$ is given by the sum of orbital and spin momentum ($J= \ell + s$), and in another state in which $\ell$ and $s$ couple antiparallel ($J = \ell - s$). As an example, a 3$d$ core state ($\ell = 2$)  will split in 3$d$ ($J=5/2$) and 3$d$* ($J= 3/2$) states, with the energy of the former higher than the one of the latter.}

As mentioned above, the integration over the BZ is carried out by {\tt JOINT} with the input file ({\tt case.injoint}) described in Table \ref{tab:WienWrapJn}

\begin{table}[hb]
\caption{A typical input file {\tt case.injoint} for computing the dichroic spectra. The values are the ones used for the calculation of the Ce L$_{2,3}$ edge, shown in Sec. \ref{cefe2}.}
\vskip 0.2cm
\begin{tabular}{l l l } \hline
\hline
1  \quad 9999 \quad 9999 & {\tt NYMIN}, {\tt NYMAX}, {\tt NYOCC}\\
 -1.0 \quad 0.001 \quad 2.0 &  {\tt EMIN}, {\tt DE}, {\tt EMAX}\\
 ryd & {\tt ECV}\\
 1 & {\tt XMCD}\\
 -412.2 \quad -445.0 & {\tt CORE\_ENE1}, {\tt CORE\_ENE2} \\
 3.4 \quad 3.2 & {\tt CORE1BRD}, {\tt CORE2BRD}\\
 0.1 & {\tt SPECTR}\\
\hline
\end{tabular}
\label{tab:WienWrapJn}
\end{table}

The meaning of the input parameters is the following:

\begin{itemize}
\item {\tt NYMIN}, {\tt NYMAX} are the lower and upper band index for the summation (Eq. \ref{eqpm2}). Optionally, one can provide the index of the highest occupied band  {\tt NYOCC} (known from the ground-state calculation) to avoid summation over terms (of zero value) between occupied states.
\item {\tt EMIN} and {\tt EMAX} define the energy window (in Ry) with respect to the Fermi level, for which the spectra should be printed; {\tt DE} is the corresponding increment for this energy mesh.
\item {\tt ECV} defines the units for the output (case sensitive): 'eV', 'ryd', or 'cm-1' for eV, Ry, and wavenumbers (cm$^{-1}$), respectively.
\item {\tt XMCD} is the parameter to distinguish between an XMCD calculation ({\tt XMCD = 1}) or a calculation of optical properties ({\tt XMCD = 0}).
\item {\tt CORE\_ENE1} and {\tt CORE\_ENE2} are the energies of the higher  (1s, 2s, 2p, 3s, 3p, or 3d) and lower (2p*, 3p*, or 3d*) core states, respectively. In the case of $K_{1}$, $L_{1}$, and $M_{1}$, {\tt CORE\_ENE2} is not read.
\item {\tt CORE1BRD} and {\tt CORE2BRD} are the core-hole lifetime broadenings in eV (\emph{i.e.}, $\Gamma_{c}$ in Eq. \ref{eq_corebroad}) of {\tt CORE\_ENE1} and {\tt CORE\_ENE2}, respectively. Again, in case of $K_{1}$, $L_{1}$, or $M_{1}$ {\tt core2brd} is not read.
\item {\tt SPECTR} is the spectrometer broadening.
\end{itemize}

The momentum matrix elements are read by the {\tt JOINT} routine {\tt readopx.f} from the files {\tt case.symmat1up} and/or {\tt case.symmat2up} (depending on the edge), and then the BZ integration is performed by the main routine {\tt joint.f} and the subroutine {\tt arbdosx.f}. Broadened and unbroadened dichroic and absorption spectra are then written into {\tt case.xmcd} and {\tt case.rawxmcd}, respectively. To give the user the choice to apply sum rules (see Sec. \ref{sumrules}), broadened and unbroadened (raw) spectra for the single edges are written to the files {\tt case.broad2p}, {\tt case.broad2ps}, {\tt case.raw2p}, and {\tt case.raw2ps}, where {\tt 2p} refers to the upper edge and {\tt 2ps} to the lower one.

\section{Results}
\label{results}

XMCD spectra can be calculated for any kind of system for which a Wien2k ground-state calculation has been performed, such as bulk materials or supercell calculations. In Refs. \cite{schatt2011,pardini2010} results for bulk iron, cobalt, and nickel are reported and compared to experimental as well as theoretical results from the literature. Overall good agreement was found in all cases. Here, we show two case studies: an example of a transition from $d$-like core states to $f$-like valence bands, in particular the $M_{4,5}$ edge of cerium in CeFe$_{2}$, and a dichroic excitation at the $L_{2,3}$ cerium edge of
the same system.

\subsection{$CeFe_{2}$}
\label{cefe2}

This system has been extensively investigated both theoretically \cite{eriksson1988,khowash1991,trygg1994} and experimentally \cite{paolasini1998,paolasini2003,delobbe1998,giorgetti1993}, where the main interest was focused on its anomalous physical properties in comparison with other rare-earth (R) RFe$_{2}$ compounds. In particular, it exhibits very low magnetic moments at iron and cerium sites (with experimental values of 1.40 $\mu_{B}$ and 0.70 $\mu_{B}$, respectively) which are antiferromagnetically coupled to each other.

$CeFe_{2}$ crystallizes in the cubic Fd{\=3}m structure with a lattice constant a=7.31\AA. The band structure calculation has been performed by sampling the whole Brillouin zone with 2000 k-points (163 in the irreducible part). Exchange and correlation effects are accounted for by the generalized gradient local approximation (GGA) in the version proposed by Perdew \emph{et. al.} \cite{perdew1996}. Spin-orbit coupling in the valence shell is included via a second variational scheme, as implemented in Wien2k.

\begin{figure}[h]
 \begin{center}
 \bigskip
  \includegraphics[width=\textwidth,angle=0]{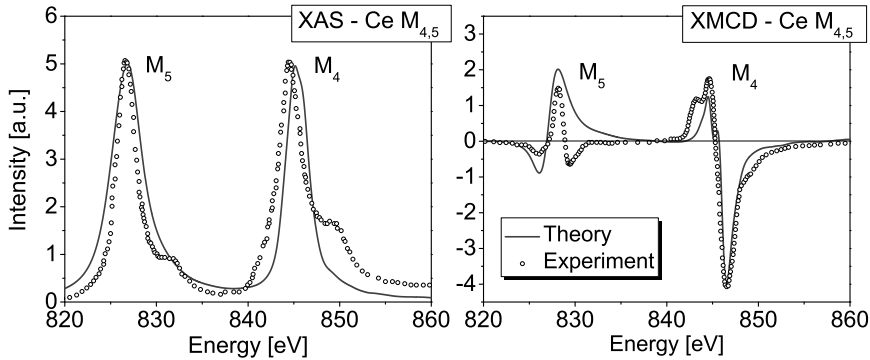}
%  \bigskip
%  \includegraphics[width=6cm,angle=0]{XMCDCeM45.eps}
  \caption{\label{CeM45} Absorption (left) and dichroic (right) spectra for the $M_{4,5}$ edge of cerium.}
 \end{center}
\end{figure}

In Fig. \ref{CeM45}, absorption and dichroic spectra are plotted for the $M_{4,5}$
edge of cerium. The theoretical absorption spectra (full line) present a very simple
line shape at both the M$_5$ (826 eV) and M$_4$ (845 eV) edges,
without any feature, whereas the experimental ones (dotted line) show
a characteristic second peak at about 5 eV above the principal ones.
This satellite has been discussed in the literature \cite{delobbe1998} and can be reproduced when including multiplet effects in the calculation. We will not focus on this point here, since it is outside
the scope of this work. Lorentzian broadenings of 1.4 eV and 0.6 eV have been chosen for the $M_{5}$ and $M_{4}$ edges, respectively,  in order to reproduce the experimental branching ratio at the two absorption edges. This choice leads to a good agreement also for the XMCD spectrum (right panel of the figure), apart from a small negative peak in the experimental $M_{5}$ edge at about 830 eV, related to the satellite, which is not reproduced in this calculation.
Applying the sum rules, we have calculated the contribution of the $f$ states to the total spin and orbital moment to be $\mu_{spin}$= -0.46 $\mu_{B}$ and $\mu_{orb}$= -0.03 $\mu_{B}$, respectively.

\begin{figure}[h]
 \begin{center}
 \bigskip
  \includegraphics[width=\textwidth,angle=0]{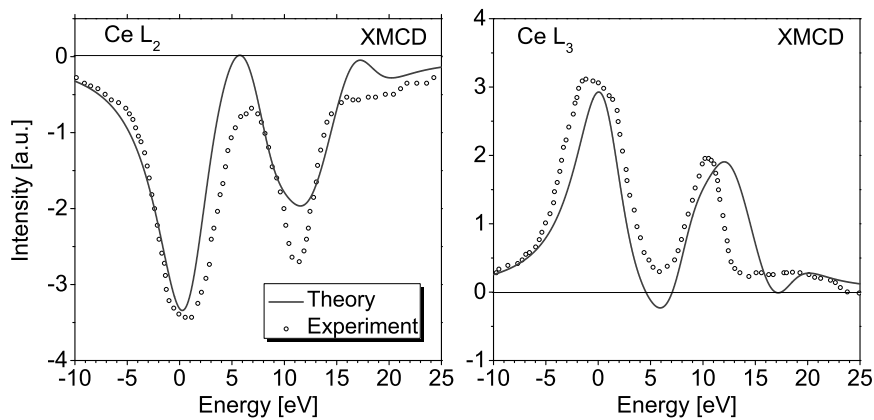}
\caption{\label{CeL23} XMCD spectra of the $L_{2}$ (left) and the $L_{3}$ (right) edge of chromium. The $L_{2}$ ($L_{3}$) edge has been convoluted
with a 3.2 eV (3.4 eV) full width half maximum Lorentzian.}
 \end{center}
\end{figure}

As an example of a transition from $p$ core states to $d$ valence bands, we report in Fig. \ref{CeL23} Ce $L_{2}$ and $L_{3}$ XMCD spectra. Again, they are in very good agreement with the experimental ones. In this case, the two edges have to be depicted in two separate windows, because the spin-orbit separation between the two core states is very strong, {\emph i.e.}, about 32.8 Ry.

Applying sum rules to the $L_{2,3}$ spectra we found a contribution from the cerium $d$ states to the total spin and orbital moment of cerium of -0.03  $\mu_{B}$ and 0.00 $\mu_{B}$, respectively. We obtain a total magnetic moment of -0.52 $\mu_{B}$ at the cerium site. This value is slightly smaller than the one (-0.69 $\mu_{B}$) found by Antonov \emph{et al.} \cite{antonov2008} with an atomic sphere approximation fully relativistic linear muffin-tin orbital (ASA-LMTO) code. In particular, they reported almost equal contributions from Ce $f$ and Ce $d$ to the total magnetic moment of Ce, whereas in the present work the Ce-$f$ contribution dominates the total moment. This small discrepancy can be explained by the different band-structure method and exchange-correlation potential adopted. As the expectation values of the moments depend in both methods on the atomic-sphere radii used, a slight difference may arise from this fact.

In order to compute the total magnetic moment of the system, we have applied sum rules to the $L_{2,3}$ spectra of iron (not shown here), finding 0.04 $\mu_{B}$ and 1.25 $\mu_{B}$ for orbital and spin moments, respectively. By summing up all moments at the iron and chromium sites ($\mu_{tot}$=$\mu_{orb}$(Ce)+$\mu_{spin}$(Ce)+2[$\mu_{orb}$(Fe)+$\mu_{spin}$(Fe)]), we obtain a value of 2.08 $\mu_{B}$, which is about 9$\%$ smaller than the experimental  value of 2.30 $\mu_{B}$.

Another point worth to mention is that from the inferred moments of Fe (1.29 $\mu_{B}$) and Ce (-0.52 $\mu_{B}$), this calculation correctly predicts an antiparallel alignment between Fe and Ce, the same as found in experiments\cite{delobbe1998} as well as in earlier calculations\cite{eriksson1988}.

\section{Conclusions}
\label{Conclusions}

We have introduced a tool for the calculation of x-ray absorption spectra at the K-, L-, and M-edges for left and right circularly polarized light in Wien2k code, by generalizing the already implemented package \emph{OPTIC}. As an example we have shown XMCD and XAS spectra of CeFe$_{2}$ at the cerium $M_{4,5}$ and L$_{2,3}$ absorption edges. The results exhibit good agreement between theoretical and experimental spectra and demonstrate the capability of the code to reliably treat such core-level excitations.

\section{Acknowledgements}

We appreciate partial support from the European Theoretical Spectroscopy
Facility (ETSF), and from the italian Ministero per l'Istruzione, l'Universit\`a e
la Ricerca (MIUR) through project PRIN/COFIN, contract 2008NX9Y7.

\end{document}